\documentclass[preprint, 12pt, final]{elsarticle}

\usepackage{epsfig}
\usepackage{subfig}
\usepackage{graphicx}

\usepackage[hidelinks]{hyperref}
\usepackage{url}
\usepackage[utf8]{inputenc}
\DeclareUnicodeCharacter{00211B}{\ensuremath{\mathcal{R}}}

\usepackage{amssymb}
\usepackage{amsthm}
\usepackage{amsmath}

\theoremstyle{plain}
\newtheorem{theorem}{Theorem}[section]

\theoremstyle{definition}

\usepackage{booktabs} 
\usepackage{colortbl}
\usepackage[table]{xcolor}

\usepackage{tcolorbox}

\usepackage{lineno,hyperref}

\begin{document}

\begin{frontmatter}

\title{Modeling the COVID-19 pandemic: A primer and overview of mathematical epidemiology}

\author[address1]{Fernando Salda\~na}
\ead{fernando.saldana@im.unam.mx}

\author[address1]{Jorge X. Velasco-Hern\'andez}
\ead{jx.velasco@im.unam.mx}

\address[address1]{Instituto de Matem\'aticas, Campus Juriquilla, 76230, Universidad Nacional Aut\'onoma de M\'exico, Qu\'eretaro, Mexico}

\begin{abstract}
Since the start of the still ongoing COVID-19 pandemic, there have been many modeling efforts to assess several
issues of importance to public health. In this work, we review the theory behind some important mathematical models that have been used to answer questions raised by the development of the pandemic. We start revisiting the basic properties of simple Kermack-McKendrick type models. Then, we discuss extensions of such models and important epidemiological quantities applied to investigate the role of heterogeneity in disease transmission e.g. mixing functions and superspreading events, the impact of non-pharmaceutical interventions in the control of the pandemic, vaccine deployment, herd-immunity, viral evolution and the possibility of vaccine escape.
From the perspective of mathematical epidemiology, we highlight the  important properties, findings, and, of course, deficiencies, that all these models have.

\end{abstract}


\end{frontmatter}

\section{Introduction}
According to the World Health Organization (WHO), the COVID-19 pandemic has caused a dramatic loss of human life and presents an unprecedented challenge to public health. The pandemic has also disrupted the global economy, the food system, education, employment, tourism, and several other aspects of life. In response to the COVID-19 crisis, the scientific community has acted fast to better understand the epidemiological, biological, immunological, and virological aspects of the SARS-CoV-2. Mathematical models have played a significant role to support public health preparedness and response efforts against the ongoing COVID-19 pandemic \cite{acuna2020modeling, aguiar2020modelling, aleta2020modelling, althouse2020superspreading, angulo2020simple, castro2021prioritizing,  contreras2020real, day2020evolutionary, de2021se, endo2020estimating, flaxman2020estimating, fontanet2021sars, ganyani2020estimating, goldstein2021vaccinating, kain2021chopping, kochanczyk2020super, mena2020using, saad2021epidemiological, saldana2020modeling, santamaria2020possible, santana2020lifting}. From the start of the pandemic, modelers have attempted to forecast the spread of COVID-19 in terms of the expected number of infections, deaths,  hospital beds, intensive care units, and other health-care resources. However, although models are useful in many ways, their predictions are based on a set of hypotheses both mathematical and epidemiological for two complex evolving biological entities, populations of hosts and pathogens. Therefore, simulation models are far from perfect and their forecasts, predictions, and scenarios should be taken cautiously \cite{eker2020validity, thompson2020key}.  There is a group of models that have played a significant role in the COVID-19 epidemic. This group is the one of the so-called compartmental models, either deterministic or stochastic, that subdivide a given human population into sets of individuals distinguished by their disease status. So we can, for example, have susceptible, infected, recovered, immune hosts where each of these classes can be further subdivided as necessary. In the COVID-19 pandemic it has been useful, for example, to consider several classes of infectious individuals as are confirmed, asymptomatic infected, symptomatically infected, isolated, and so on. Infection occurs when an infectious individual enters into contact with a susceptible one and pathogen transmission ensues. In infectious disease models, perhaps the most important component is the one that describes this infection process, the so-called mixing function. The mixing function introduces one of the basic properties of epidemic systems: heterogeneity. This heterogeneity is expressed in the different ages, tastes, activities, and other sexual, behavioral, social, genetic, and physiological traits that define an individual and a group of individuals in a population. We do not mix randomly because we tend to mix and interact with people that are like us in some particular, specific way.
The evolution of the present pandemic has been driven mainly by heterogeneity. All around the World, mitigation measures were implemented to control disease spread. The level of enforcement and compliance of such measures, however, widely varied across the globe. Moreover, at times when mitigation measures were partially relaxed, flare-ups and secondary or tertiary outbreaks have occurred.  Many have been associated with so-called superspreading events. These are characterized as periods of time where large numbers of individuals congregate in close contact increasing the average transmission. These events then become foci of infections when individuals return to their homes or communities and start, by doing so, a new wave of local disease transmission. The understanding and modeling of superspreading events constitute a present challenge of significant importance for the control of disease spread \cite{althouse2020superspreading, endo2020estimating, hebert2020beyond, liu2020secondary, kain2021chopping, kochanczyk2020super}.

At the beginning of the pandemic, given the lack of  treatments or effective vaccines, we have relied on the implementation of non-pharmaceutical interventions to prevent disease spread. These types of measures can be rather effective in diminishing transmission but largely rely on individual customs, beliefs, and education. Vaccines on the other hand, available since the beginning of 2021, present problems of their own. One is the interplay between coverage, efficacy, and design. The vaccines that are in the market as this work is being written, are designed for the viral variants prevalent in the first six months of the epidemic. Now we have several new variants (that, curiously, arose in the countries where the epidemic has been worse: UK, USA, Brazil) that have higher transmissibility and the same type of mutations, facts that present worrisome perspectives regarding the possibility of virus variants evolving to escape the action of vaccine. Coverage is a problem too since there has been limited vaccine supply. Without sufficient coverage, the epidemic will linger for many more months with the consequent burden on the economy and public health of many countries. Mathematical tools are being used to provide criteria for the deployment and vaccine roll-out and the impact of these on epidemic evolution.
Finally, a  problem of interest and where modeling is necessary is the analysis of syndemic diseases, that is, diseases that co-circulate in the same region, time, and populations. As an example, we have the interaction between influenza and COVID-19 that so far has resulted rather benign since the mitigation measures that have only reduced the prevalence of SARS-CoV-2, have practically eliminated influenza in the season October 2020-March 2021 in the Northern hemisphere \cite{zegarra2021co}.

This work has as the main objective to review particular models that have been or are being used to address the list of problems commented upon in this Introduction. We will highlight the most important properties, findings, and, of course, deficiencies, that all these models have. The perspective is that of two mathematical epidemiologists involved in the Public Health application of models. We hope the approach and perspective will be of interest to a general mathematical audience.

\section{Kermack-McKendrick models and key epidemiological parameters}
The classical work of Kermack \& McKendrick published in 1927 \cite{kermack1927contribution} is a milestone in the mathematical modeling of infectious diseases. The model introduced in \cite{kermack1927contribution} is a complex age of infection model that involves integro-differential equations and the now very famous compartmental SIR model as a special case:
\begin{equation}
\begin{aligned}
\dot{S} &= -\beta SI,\\
\dot{I} &= \beta SI - \gamma I,\\
\dot{R} &= \gamma I.\\
\end{aligned}
\label{sir}
\end{equation}
The variables $S$, $I$, $R$, in system represent the  number of susceptible, infectious and recovered individuals in the constant total population $N=S+I+R$. The parameter $\gamma$ is the recovery rate and $\beta=c\phi$, is the effective contact rate which is product of the number of contacts $c$ and the probability of infection given a contact.

A well known result for system \eqref{sir} is that the mean infectious period is
\begin{equation}
\int_{0}^{\infty}\gamma t\exp(-\gamma t)dt=\dfrac{1}{\gamma}\label{inf_period}
\end{equation}         
and therefore the residence times in the infectious state are exponentially distributed. In other words, the probability of recovery per unit of time is constant, regardless of the time elapsed since infection \cite{vergu2010impact}. Using \eqref{inf_period}, we can write the basic reproduction number as
\begin{equation}
\mathcal{R}_{0}=\dfrac{\beta}{\gamma}S_{0}\label{r0_s0}.
\end{equation}
$\mathcal{R}_{0}$ measures the expected number of infections generated by a single (and typical) infected individual during his/her entire infectious period ($1/\gamma$) in a population where all individuals are susceptible to infection ($\beta S_{0}=\beta N$). At the early phase of the outbreak,
\begin{equation}
\dot{I}=\gamma (\mathcal{R}_{0} -1)I,\quad \text{and}\quad I(t)=I(0)\exp\left( \gamma(\mathcal{R}_{0}-1)t\right) 
\end{equation}
so the infectious class grows initially if $R_{0}>1$ and goes to zero otherwise. Furthermore, $r=\gamma(\mathcal{R}_{0}-1)$ is the grow rate of the epidemic and 
\begin{equation}
\mathcal{R}_{0}=1+\dfrac{r}{\gamma}\label{r0_sir}.
\end{equation}
Observe that formula \eqref{r0_sir} provides a natural method to compute $\mathcal{R}_{0}$ without the need of estimate the initial susceptible population $S_0$.
Another fundamental property of the SIR model \eqref{sir} is that 
\begin{equation}
\lim_{t\rightarrow \infty}S(t)=\lim_{t\rightarrow \infty}S(0)\exp\left( -\beta\int_{0}^{t}I(s)ds \right)  =S_{\infty}>0,\quad \lim_{t\rightarrow \infty}I(t)=0
\end{equation}
so the outbreak will end leaving susceptible individuals who escape infection. Moreover, direct computations allow us to obtain a relationship between the total number of cases $S(0)-S_{\infty}$ and the cumulative number of infections at time $t$, denoted $C(t)$, 
\begin{equation}
C(t) = -(\log S(t)-\log S_{0}) = \dfrac{\gamma}{\beta}\int_{0}^{t}I(s)ds
\end{equation}
and 
\begin{equation}
C(\infty)=\lim_{t\rightarrow\infty}C(t)=\log\dfrac{S_{0}}{S_\infty} = \dfrac{\gamma}{\beta}\int_{0}^{\infty}I(s)ds
\end{equation}
Then we obtain the final size equation
\begin{equation}
S(0)\exp\left( -\mathcal{R}_{0}C(\infty)\right) = S(0)+I(0)-\dfrac{1}{\mathcal{R}_{0}}C(\infty)
\end{equation}
that allow us to estimate the expected number of total cases $C(\infty)$ given $\mathcal{R}_{0}$ and the size of the initial susceptible population.

It is important to remark that compartmental Kermack-McKendric-type models rely on specific assumptions that should be taken with care when models are applied to real-life problems such as modeling the COVID-19 pandemic. For example, from \eqref{inf_period} we deduce that the SIR model assumes that the time during which an infectious individual can transmit the disease is constant and equal to the recovery period. This is an acceptable approximation for some diseases but, for others, especially those whose recovery time is long, it is not. In general, the capacity of infectious individuals to infect another person will depend on the age of infection \cite{kermack1927contribution}. Let $i(a,t)da$ be the density of infectious individuals at time $t$ with an age of infection between $a$ and $a+da$, the evolution of such population is governed by the McKendrick-von Foester equation
\begin{equation}
\dfrac{\partial}{\partial t}i(a,t) + \dfrac{\partial}{\partial a}i(a,t)+\gamma(a)i(a,t)=0, \label{MvonF}
\end{equation}
where the recovery rate now depends on the age of infection $a$. The boundary and initial conditions are
\begin{equation}
i(a,0)=\phi(a), \quad i(0,t)=B(t)=\int_{0}^{\infty}\beta(a)i(a,t)da,
\end{equation}
where $\beta(a)$ is the effective contact rate as function of the age of infection and $\phi(a)$ is the age of infection initial distribution. For the simple case in which the solution of \eqref{MvonF} is separable i.e. $i(a,t)=T(t)A(t)$,  one can see that \eqref{MvonF} has a solution if there is a unique $r\in \mathbb{R}$ that satisfies the characteristic equation 
\begin{equation}
\int_{0}^{\infty}e^{-ra}\beta(a)e^{\left( -\int_{0}^{a}\gamma(s)ds\right)} da=1.\label{characteris}
\end{equation} 

The parameter $r$ now is interpreted as the intrinsic growth rate of the epidemic and is directly related with the epidemic doubling time, the amount of time in which the cumulative incidence doubles, $T_{d}=\log 2/r\approx 1/r$. If there is an increase in the doubling time then the transmission is decreasing. To obtain the basic reproduction number, we need to look at the expression
\begin{equation}
K(a)=\beta(a)e^{-\int_{0}^{a}\gamma(s)ds}
\end{equation}
which is the product of the contact rate and the probability of remain infectious at the age of infection $a$. Hence, $K(a)$ gives the number of secondary infections generated by an individuals of age of infection $a$ and
\begin{equation}
\mathcal{R}_{0}=\int_{0}^{\infty}K(a)da.
\end{equation}
In the SIR model \eqref{sir}, the probability of being infectious at time $t$ is exponential, and the transmission rate at the beginning of the epidemic $\beta S_0$ is independent from the age of infection, thus
\begin{equation}
\mathcal{R}_{0}=\int_{0}^{\infty}K(a)da = \beta S_0 \int_{0}^{\infty}e^{-\gamma t}dt = \dfrac{\beta}{\gamma}S_{0}
\end{equation}
and we recover the expression \eqref{r0_sir}.

The function $K(a)$ has also been called the distribution of the generational interval. This is one of the fundamental concepts for the computation of $\mathcal{R}_{0}$. However, it has been largely neglected in the mathematical community due to the almost exclusive attention to the computation of reproduction numbers based on compartmental epidemic models. Estimates of the reproduction number together with the generation interval distribution can provide insight into the speed with which a disease will spread as shown in \cite{ganyani2020estimating} for the case of COVID-19. The density of the generational interval is defined as
\begin{equation}
G(a)=\dfrac{K(a)}{\mathcal{R}_{0}},
\end{equation}
clearly, $0\leq G(a)\leq 1$ for all $a$. Function $G(a)$ measures the time between the infection of a primary case and one of its secondary cases \cite{kenah2008generation, wallinga2007generation}. In other words, the generational interval $G(a)$ is the age of infection that separates the infector from the infectee. From \eqref{characteris} and the definition of $G(a)$, we obtain that the inverse of the $\mathcal{R}_{0}$ is the result of the Laplace transform, or moment-generating function, of the distribution of the generational interval evaluated at the intrinsic grow rate
\begin{equation}
M(r)=\int_{0}^{\infty}e^{-ra}G(a)da=\dfrac{1}{\mathcal{R}_{0}}.\label{M}
\end{equation}

The generation interval of an epidemic outbreak is calculated directly from incidence data, using the symptom onset date as an approximation to the contagion date, and making extensive use of screening and contact tracing methods \cite{ganyani2020estimating, nishiura2020serial, rai2020estimates}. This approximation is known as the serial interval. We must remark that the mean of the distribution for the serial or generational interval does not necessarily coincide with the infectious period which is only an approximation particular to Kermack-McKendrick type models. Using \eqref{M}, we can calculate $\mathcal{R}_{0}$ for different distributions for the serial interval \cite{wallinga2007generation}. For example, let $T_g$ be the mean serial interval, a straightforward computation allow us to see that for the exponential distribution $G(a)=\gamma \exp(-\gamma t)$ we obtain
\begin{equation}
\mathcal{R}_{0}=\dfrac{1}{M(r)}=1+\dfrac{r}{\gamma}
\end{equation}
where $T_g = 1/\gamma$ and again we deduced \eqref{r0_sir}. For a normal distribution we have
\begin{equation}
G(a)=\dfrac{1}{\sqrt{2\pi}\sigma}e^{(a-1/\gamma)^{2}/4\sigma^{2}},\quad \mathcal{R}_{0}=\dfrac{1}{M(r)}= e^{r/\gamma-\sigma^{2}r^{2}/2}
\end{equation}
so, in this case, an increase in the variance decreases the value of the basic reproduction number. For new emerging diseases as COVID-19, the natural way to obtain $\mathcal{R}_{0}$ is from the observed initial growth rate $r$ of the epidemic. However, as we have shown, the equation relating these two parameters varies concerning the distribution of the infection period so one must be careful since the distribution of the generation interval must be known before we can apply a given relationship for the infection understudy \cite{gostic2020practical, wallinga2007generation}.

\section{Mixing functions and disease spread}
The SIR model \eqref{sir} is one of the simplest that can be written for a communicable disease. As explained before, it makes the overly simplifying assumption that the mass action law, expressed in the product $SI$, is a good approximation of the mixing of the population that results in infectious contacts. However, this hypothesis is only a very rough approximation to what really happens. To illustrate this consider the spread of a $SI$ disease in a population subdivided into $n$ disjoint groups first presented in \cite{LajYork1976}. Each group is homogeneous in itself meaning that the recovery rates are the same for all the individuals belonging to that group and their contacts between individuals are function only of the group they belong to. Define as $\beta_{ij}$ the effective contact rate of the susceptible individuals in the group $i$ with infectious individuals in the group $j$, $N_i$ is the total population size of the group $i$, and $\gamma_i$ is the recovery rate of infected individuals in the group $i$. Another hypothesis is that $\beta_{ij}=\beta_{ji}$, so contacts are symmetric although this may not be always true. We will show a more general condition on the contacts rates later in this section. Note that since $S_i=N_i-I_i$ we need only write the equations for the infected individuals in each group $i$, giving
\begin{align}\label{eq:lj}
    \dot{I}_i&=-\gamma_i I_{i}+\sum_{j=1}^n\beta_{ij}N_iI_j-\sum_{j=1}^n\beta_{ji}I_iI_j,
\end{align}
which is studied in the set $\Omega=\Pi_{i=1}^n[0,N_i]$. 
This system has a globally asymptotically stable disease-free equilibrium at $I=(I_1,\cdots I_n)=0$ and another endemic equilibrium, with $I=k$ with $k$ a constant vector, that is globally asymptotically stable when the disease-free equilibrium loses its stability. We form the matrix $A=(a_{ij})$ where $a_{ij}=\beta_{ij}$ when $i\neq j$, and $a_{ij}=\beta_{ij}-\gamma_i$ if $i=j$ and define the column vector  $B(I)$ with components $B_i=-\sum_{j=1}^n\beta_{ji}I_iI_j$, then equation \eqref{eq:lj} can be rewritten \cite{LajYork1976} as
\begin{align}
  \dot{I}_i&=A I+B(I). \label{eq:coop}
\end{align}
The main result of  \cite{LajYork1976} is the following:
\begin{theorem}\label{thm:LJ}
Consider the system given by \eqref{eq:coop} where $A$ is an $n\times n$ irreducible matrix and $B$ is continuously differentiable in a region $D$ of $R^n$. Assume that
\begin{itemize}
    \item the compact convex set $C\subset D$ contains the origin and is positively invariant for (\ref{eq:coop}),
    \item $\lim_{I\to 0}\frac{||B(I)||}{||I||}=0$,
    \item There exists $\lambda > 0$ and a real eigenvector $v$ of $A^T$ such that $v\cdot v\geq \lambda ||I||$ for all $I\in C$,
    \item $v\cdot B(I)\leq 0$ for all $I\in C$,
    \item Either $I=0$ is globally asymptotically stable in $C$, or for any $I_0\in C-{0}$ the solution $\phi(t,I_0)$ of (\ref{eq:coop}) satisfies $\lim_{t\to\infty}||\phi(t,I_0)||\geq m$, $m$ independent of $I_0$. Lastly, there exist a constant solution of (\ref{eq:coop}) $I=k$, $k\in C-{0}$.
    \end{itemize}
\end{theorem}
In biological terms, this result simply states that if the basic reproductive number $\mathcal{R}_0<1$, then the disease-free equilibrium is unique and asymptotically stable, but if $\mathcal{R}_0>1$, then it is unstable and there exists another equilibrium, the endemic equilibrium, that is globally asymptotically stable.  The Lajmanovich and York model as the system (\ref{eq:lj}) is known, was one of the first multigroup models that shows that its dynamics can be fully characterized in terms of the fundamental parameter $\mathcal{R}_0$ and the transcritical bifurcation that the equilibria suffer at the critical value $\mathcal{R}_0=1$ \cite{Hethcote1999}. Several studies have shown, however, that epidemiological models that follow the Lajmanovich and York characterization are not that general. They are limited in the hypothesis regarding the total population, which is assumed constant, and the mixing pattern among population subgroups. The existence of subthreshold endemic states, that is, endemic states that exist even when $\mathcal{R}_0<1$, have been shown for directly transmitted diseases and vector transmitted diseases (see, for example, \cite{Dushoff1998, gumel2012causes, KribsVH2000, Garba2008, saldana2019role, villavicencio2008latency}). A simple example of this type of behavior is illustrated in a generalization of the model first published in \cite{KribsVH2000} that is pertinent for the study of vaccination policies during the present pandemic. A close variant of this model is briefly presented now. Let $S$, $I$, $V$, and $R$ denote the subpopulations of susceptible, infectious, vaccinated, and immune individuals respectively. Let $\beta$ be the effective contact rate, $\sigma$ the proportion of reduction in the contact rate due to the vaccine, $\omega$ the waning rate of both vaccine and natural immunity, $\gamma$ the recovery rate, and $\Lambda$, $\mu$ the equal birth and mortality rates, respectively. The model stands

\begin{equation}
\begin{aligned}
\dot{S}&=\Lambda -\beta S \frac{I}{N} -(\mu +\phi)S+\omega V,\\
\dot{V}&=\phi V-(1-\sigma) V\frac{I}{N}-(\mu+\omega)V,\\
\dot{I}&=\beta \frac{I}{N}(S+(1-\sigma) V)-(\gamma+\mu)I,\\
\dot{R}&=\gamma I- (\omega+\mu)R.
\end{aligned}\label{kribsVH}
\end{equation}

\begin{figure}[hbtp]
 \centering{
    \includegraphics[width=0.5\textwidth]{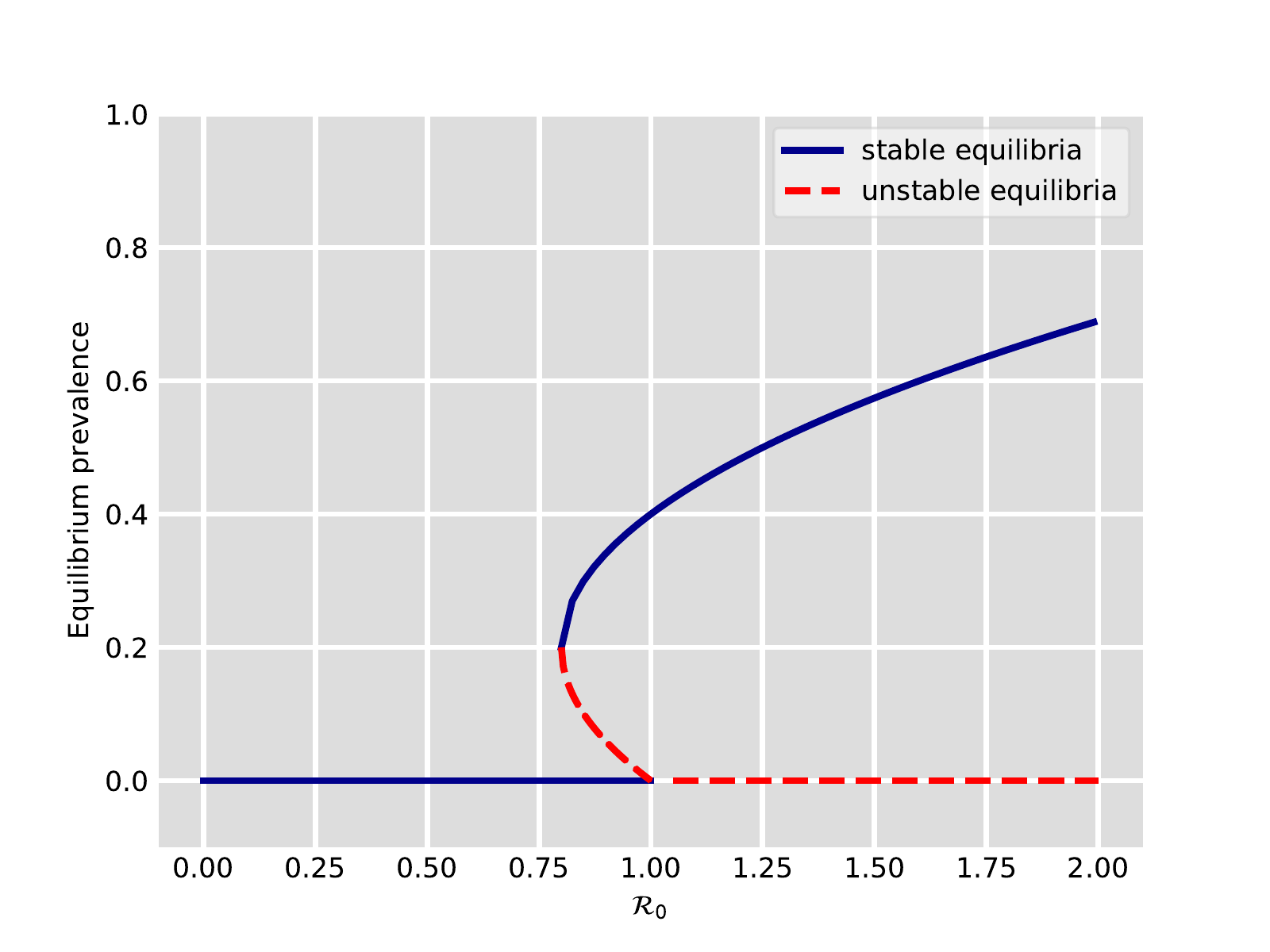}
    }
 \caption{Schematic representation of a typical backward bifurcation in epidemic models. The basic reproduction number $\mathcal{R}_0$ is showed in the horizontal axis and the vertical axis is the prevalence of the infection (fraction) at the equilibrium. Solid blue lines represent stable equilibria and dotted red lines represent unstable equilibria.}
 \label{fig:bb}
\end{figure}

Using model \eqref{kribsVH}, one can show that, besides the well-known forward bifurcation in which $\mathcal{R}_{0}<1$ is a necessary and sufficient condition for disease elimination, epidemic models may present a backward bifurcation where a stable endemic equilibrium (EE) co-exists with an unstable EE and a  stable disease-free equilibrium for $\mathcal{R}_{0}<1$ \cite{KribsVH2000}. In figure \ref{fig:bb} we present and schematic representation of the backward bifurcation phenomena. In terms of disease control, a backward or subcritical bifurcation implies that $\mathcal{R}_0$ has to be reduced below a value lower than one and, in some cases, difficult to estimate. Hence, a backward bifurcation is usually considered a undesirable phenomenon since control polices turn out to be more complicated \cite{barradas2019backward}.  

\par 

Another factor that alters the neat result in Theorem \ref{thm:LJ} by Lajmanovich and Yorke is the fact that the effective contact rates are not constant. They, of course, can vary when subject to climatic variability but there is a more basic reason why they are variable: effective contact rates depend on the sizes of the groups that form the population and, more importantly, on the mixing preferences of those groups. Busenberg and Castillo-Chavez \cite{Busenberg1991} after the pioneering work of \cite{Jacquez1988} constructed an axiomatic framework that clearly defines the conditions that contact rates should satisfy to be consistent with the dynamics of epidemiological processes. Following \cite{Busenberg1991}, define $c_{ij}$ the proportion of contacts that individuals from group $i$ have with those in group $j$. Then these contact rates must satisfy the following conditions to be consistent with the epidemic dynamics:
\begin{enumerate}
    \item $c_{ij}\geq 0$,
    \item $\sum_{j=1}^nc_{ij}=1$ for $j=1,...,n.$,
    \item $a_iN_ic_{ij}=a_jN_jc_{ji}$,
\end{enumerate}
where $N_i$ is the size of group $i$ and $a_i$ is the activity or risk level of group $i$. Two particular contact patterns satisfy the above axioms. The proportional mixing where contacts between groups $i$ and $j$ are separable $c_{ij}=c^a_ic^b_j$ which implies, using axioms 2 and 3 above, that $c_{ij}=Cc_j^b=\hat{c}_j$ with
$$\hat{c}_j=\frac{a_jN_j}{\sum_{i=1}^na_iN_i}.$$
Proportionate mixing refers to the fact that contacts are distributed according to the proportion of each group in the overall population.
The other important contact function is the so-called preferential mixing. Here we divide the contacts of group $i$ into two parts: a fraction of the contacts, $\epsilon_i$, is reserved for within group contacts (group $i$) and the other fraction $1-\epsilon_i$, is distributed among all other groups according to the following formula
$$c_{ij}=\epsilon_i\delta_{ij}+(1-\epsilon_i)\frac{(1-\epsilon_j)a_jN_j}{\sum_{i=1}^n(1-\epsilon_j)a_iN_i},$$
with $\delta_{ij}=0$ if $i\neq j$ and $\delta_{ij}=1$ if $i= j.$ 
Many of the contact matrices that have been evaluated in the POLYMOD study \cite{Mossong2008, polymod} show a pattern similar to that of preferential mixing characterized for being diagonally dominant since most population groups tend to share a high proportion of their contacts within the same group, in this case, defined by age classes. However, special interaction as those of children and adults which tend to have strong sub-diagonal components, and work settings where age groups are bounded above and below, do not arise from the preferential mixing formula. Glasser and coworkers \cite{Glasser2012} have generalized contact matrices and the preferential mixing function to other situations.

\section{Modeling the early phase of the COVID-19 epidemic outbreak}

An important feature of the COVID-19 disease is the incubation period, which is the time between exposure to the virus and symptom onset. The average incubation period is 5-6 days but can be as long as 14 days \cite{lauer2020incubation}. The SEIR model

\begin{equation}
\begin{aligned}
\dot{S} &= -\beta IS,\\
\dot{E} &= \beta IS- \sigma E,\\
\dot{I} &= \sigma E - \gamma I,\\
\dot{R} &= \gamma I,\\
\end{aligned}
\label{seir}
\end{equation}
allow us to incorporate the mean incubation period ($1/\sigma$) via the compartment $E$ that includes exposed individuals who had been infected but are still not infectious. The epidemic grow rate $r$ of the SEIR model \eqref{seir} can be obtained exploring the local asymptotic stability of disease-free equilibrium (DFE) $E_{\circ}=(N(0),0,0,0)$. In particular, $r$ is computed as the dominant eigenvalue of the Jacobian of system \eqref{seir} evaluated at $E_{0}$. In this case, $\mathcal{R}_{0}$ and $r$ are related as follows:
\begin{equation}
\mathcal{R}_{0}=\left(1 + \dfrac{r}{\sigma} \right) \left(1 + \dfrac{r}{\gamma} \right) 
\end{equation}

\begin{figure}[hbtp]
 \centering{
    \includegraphics[width=0.5\textwidth]{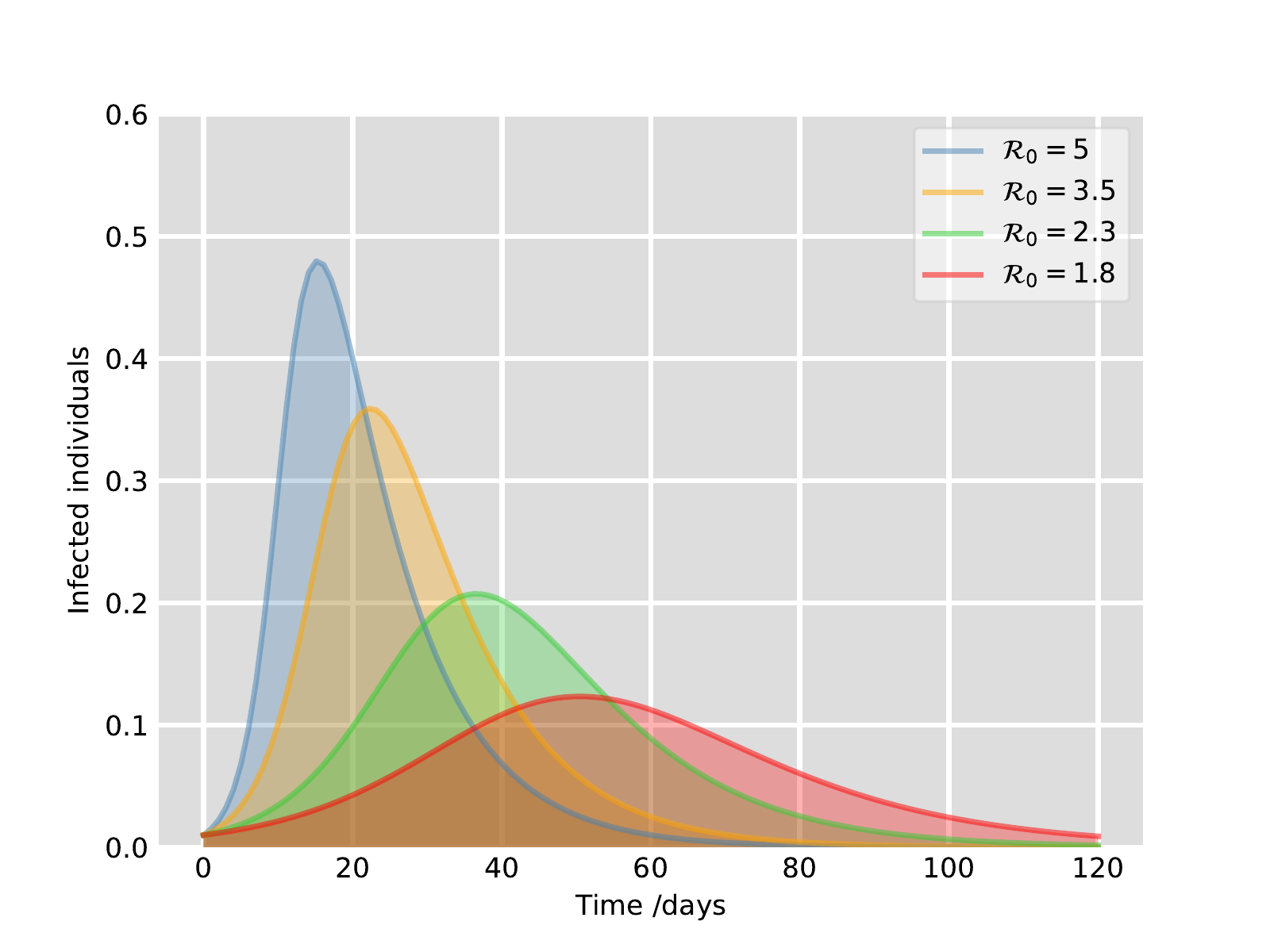}
    }
 \caption{Theoretical evolution of the epidemic curve (percentage prevalence of the infection) for different values of the basic reproduction number.}
 \label{fig:r0}
\end{figure}

The approximation for the serial interval of the SIR model is $T_g = 1/\gamma$; for the SEIR model is $T_g = 1/\gamma + 1/\sigma$. Hence, the conceptual approximation of the serial interval given by the SEIR model is more accurate in the sense that incorporates, both, the incubation and the recovery periods. To have a better understanding of how the model assumptions and data may affect the estimation of $\mathcal{R}_{0}$, let us consider the following example in the COVID-19 context. Consider an average incubation period of $1/\sigma=5$ days \cite{lauer2020incubation} and recovery time of $1/\gamma=10$ days. The epidemic doubling time at the early phase of the epidemic in Hubei Province, where SARS-CoV-2 was first recognized, has been approximated to be $T_{d}=2.5$ days \cite{muniz2020doubling} so $r=\log 2/T_{d}= 0.277$ so the SIR model estimates $\mathcal{R}_{0}=2.77$ and the SEIR model estimates $\mathcal{R}_{0}=3.8364$. On the other hand, the estimated serial interval in Hubei from contact tracing data has a median of $4.6$ days \cite{yang2020estimation}; hence, using the real serial interval the SIR model predicts $\mathcal{R}_{0}=2.27$. The overestimation of the serial interval leads to the overestimation of $\mathcal{R}_{0}$, which in turn overestimates the growth rate $r$. This implies that Kermack-McKendrick-type models may erroneously anticipate the epidemic peak time (the date of maximum incidence) and also overestimate the final epidemic size (see Fig. \ref{fig:r0}).

\par 

\begin{figure}[hbtp]
 \centering{
    \includegraphics[width=0.5\textwidth]{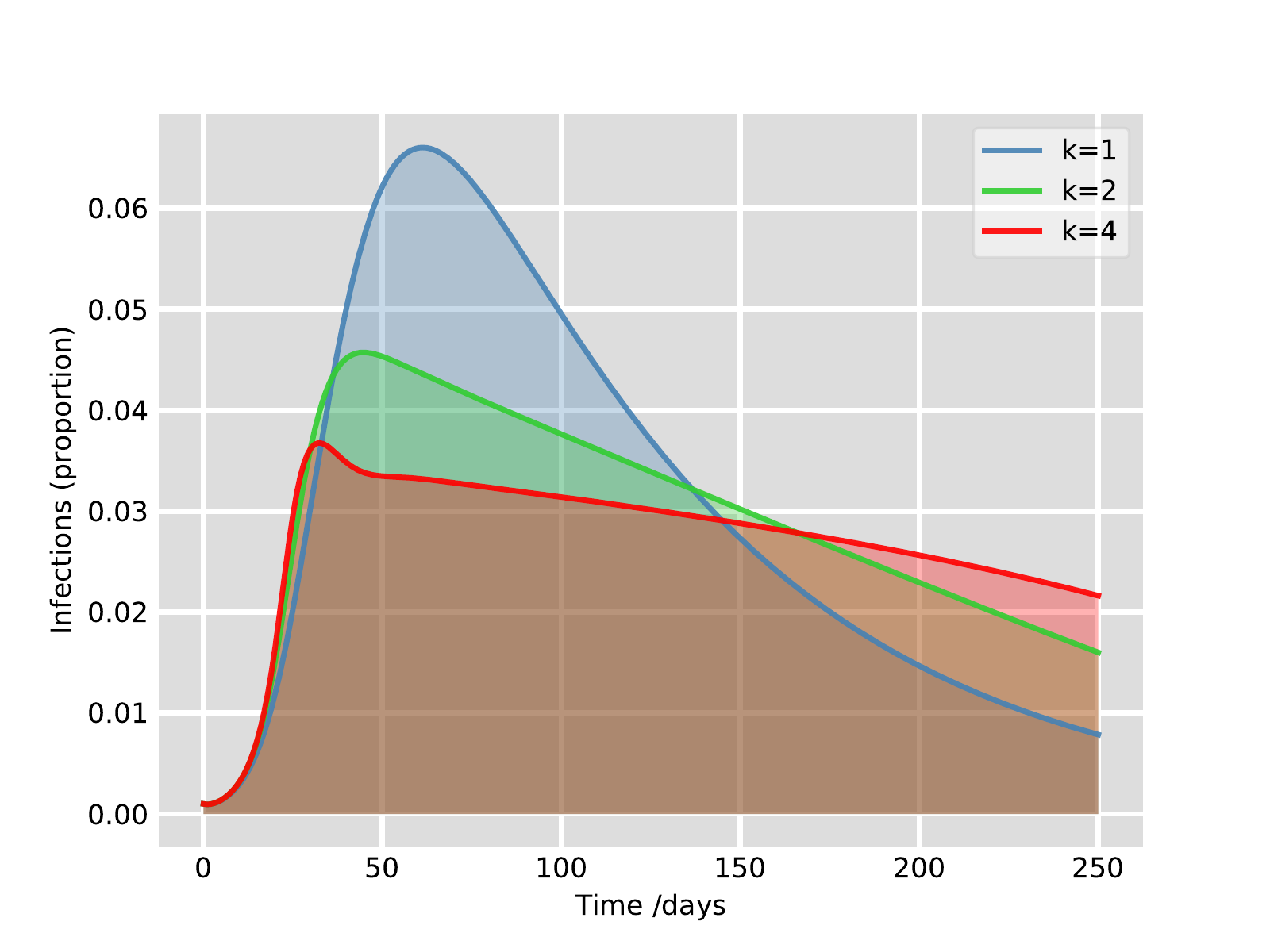}
    }
 \caption{Plateau-like states induced by death-awareness social distancing. Epidemiological parameters are fixed to mimic COVID-19 dynamics as follows: $\beta=0.5$ days$^{-1}$, $1/\sigma=5$ days, $1/\gamma =10$ days, $f=0.02$, $k$ varies and we assume $\delta_c=5\times 10^{-5}$. }
 \label{fig:plateau}
\end{figure}

Standard deterministic Kermack-McKendrick type models that neglect demographic dynamics predict a single epidemic wave (see Fig. \ref{fig:r0}); however, rather than a single peak and near-symmetric decline, the COVID-19 epidemic in several cities have shown plateau-like states followed by epidemic rebounds \cite{mena2020using, solbach2020antibody, santana2020lifting, weitz2020awareness}. This phenomenon is in part due to changes in transmission induced by lockdowns and other non-pharmaceutical interventions implemented by public health officers to reduce the epidemic burden. Besides, individuals are constantly changing their behavior and mobility patterns depending on the perceived risk of acquiring the infection, so they may reduce their contacts at times of high incidence \cite{weitz2020awareness}. How awareness-driven behavior modulates the epidemic shape has been investigated in \cite{weitz2020awareness} modifying the SEIR model \eqref{seir} as follows:

\begin{equation}
\begin{aligned}
\dot{S} &= -\dfrac{\beta IS}{1+(\delta/\delta_c)^{k}},\\
\dot{E} &= \dfrac{\beta IS}{1+(\delta/\delta_c)^{k}}- \sigma E,\\
\dot{I} &= \sigma E - \gamma I,\\
\dot{R} &= (1-f)\gamma I,\\
\dot{D} &= f\gamma I
\end{aligned}
\end{equation}
where $D$ measures the number of deaths in the population. The parameter $f$ measures the mean fraction of people who die after contracting COVID-19. The transmission rate is affected by the death-awareness social distancing rate $\delta=\dot{D}$, the half-saturation constant $\delta_c\geq 0$, and the sharpness of change in the force of infection $k\geq 1$ \cite{weitz2020awareness}. The introduction of awareness in the SEIR model \eqref{seir} renders scenarios in which a plateau-like behavior appears, that is, the number of daily infections decreases at a very slow pace after the peak  (see Fig. \ref{fig:plateau}).

\subsection{The role of asymptomatic transmission}

Another important feature of infection by SARS-CoV-2 is that some people are infected and can transmit the virus, but do not experience any symptoms \cite{park2020time}. Current evidence suggests that asymptomatic carriers will transmit the infection to fewer people than symptomatic individuals but their real contribution to transmission is difficult to estimate since they are expected to have more contacts than symptomatic carriers  \cite{asymptomaticCont}. These factors may have an impact on disease dynamics. Hence, we extend the SEIR model \eqref{seir} to illustrate the role of the asymptomatic carriers.

The model is given by the following system of non-linear differential equations:

\begin{equation}
\begin{aligned}
\dot{S} &= -(\beta_{A}A + \beta_{I}I ) S,\\
\dot{E} &= (\beta_{A}A + \beta_{I}I ) S- \sigma E,\\
\dot{A} &= (1-p)\sigma E - \gamma_{A}A,\\
\dot{I} &= p\sigma E - \gamma_{I}I,\\
\dot{R} &= \gamma_{A}A + \gamma_{I}I,\\
\end{aligned}
\label{seair_model}
\end{equation} 
where the class $A$, represents asymptomatic infectious individuals. The total population at time $t$, is now $N(t)=S(t)+E(t)+A(t)+I(t)+R(t)=1$. The parameters $\beta_{A}$ and $\beta_{I}$ represent the effective contact rates of the asymptomatic and symptomatic infectious classes, respectively. A proportion $p$ of the exposed individuals $E$ will transition to the symptomatic infectious class $I$ at a rate $\sigma$, while the other proportion $1-p$ will enter the asymptomatic infectious class $A$. The mean infectious periods in the asymptomatic and symptomatic infectious classes are $1/\gamma_{A}$, and $1/\gamma_{I}$, respectively. These individuals gain permanent immunity and move to the recovered class $R$. However, we remark that this assumption is only valid when studying the first outbreak because to date, it is still unknown how long natural immunity will last and there have been already confirmed cases of coronavirus reinfections \cite{parry2020covid, saldana2020trade, to2020serum, west2021case}.

\subsubsection{Basic properties of the SEAIR model}
A common first step in analyzing compartmental epidemic models is finding the equilibrium points. Setting the right-hand side of system \eqref{seair_model} equal to zero, we see that there is a continuum of DFE of the form 
\begin{equation}
E_{0}=(S^{*}, E^{*}, A^{*}, I^{*}, R^{*})=(N(0), 0, 0, 0, 0)\label{DFE}
\end{equation}
where $N(0)=S(0)$ is the number of susceptible individuals at the initial time. For model \eqref{seair_model}, there is no endemic equilibrium. This is because an endemic equilibrium needs a continuous supply of susceptible individuals that generally occur via births into the susceptible population or through the waning of immunity \cite{blackwood2018introduction}. However, model \eqref{seair_model} assumes permanent immunity and does not consider demographic dynamics i.e. births and deaths in the population. 

From the model equations \eqref{seair_model}, it is easy to see that $\dot{N}=\dot{S}+\dot{E}+\dot{A}+\dot{I}+\dot{R}=0$, therefore the total population is a constant $N(t)=N(0)$ for all $t$ and the solutions of system \eqref{seair_model} are bounded. The biologically feasible region is
\begin{equation*}
\Omega=\left\lbrace (S, E, A, I, R)\in \mathbb{R}_{+}^{5}: S(t)+E(t)+A(t)+I(t)+R(t)=N(t)\right\rbrace.
\end{equation*}\par 
Let $X(t)$ be the solution of system \eqref{seair_model} for a well-defined initial condition $X(0)\in \Omega$. Since $X_{i}=0$, implies $\dot{X}_{i}\geq 0$ for any state variable, then $X(t)\in \Omega$ for all $t>0$. Thus, solutions trajectories satisfy the usual positiveness and boundedness properties and the model is both epidemiologically and mathematically well posed \cite{hethcote2000mathematics}.

The local stability of the DFE is usually explored via the basic reproduction number. As we have mentioned, the mathematical expression for $\mathcal{R}_0$ depends directly on the model assumptions and structure. There are several tools for the computation of the basic reproduction number \cite{heffernan2005perspectives}. Probably, the most popular is the next-generation approach \citep{diekmann1990definition} using the method of \citep{van2002reproduction}. Under this approach, it is necessary to study the subsystem that describes the production of new infections and changes among infected individuals. The Jacobian matrix $\mathbf{J}$ of this subsystem at the DFE is decomposed as $\mathbf{J}= \mathbf{F}-\mathbf{V}$, where $\mathbf{F}$ is the transmission part and $\mathbf{V}$ describe changes in the infection status. The next-generation matrix is defined as $\mathbf{K}=\mathbf{F}\mathbf{V}^{-1}$, and $\mathcal{R}_{0}=\rho(\mathbf{K})$, where $\rho(\cdot)$ denotes spectral radius.\par 
For system \eqref{seair_model}, we obtain
\begin{equation*}
\mathbf{F}=
\left[\begin{array}{ccc}
   0  &  \beta_{A}S(0) & \beta_{I}S(0) \\
   0  &  0   &  0  \\
   0  &  0   &  0  
\end{array} \right],
\quad
\mathbf{V}=
\left[\begin{array}{ccc}
 \sigma        &  0        &  0           \\
-(1-p)\sigma   &\gamma_{A} &  0           \\
   -p\sigma    &  0        & \gamma_{I}
\end{array} \right].
\end{equation*}
Therefore, the basic reproduction number is given by
\begin{equation}
\mathcal{R}_{0}=\left( 
\dfrac{(1-p)\beta_{A}}{\gamma_{A}}+
\dfrac{p\beta_{I}}{\gamma_{I}}
\right) S(0).\label{R0}
\end{equation}

$\mathcal{R}_{0}$ is a weighted average determined by the proportion, $p$, of symptomatic infected individuals. If there are no asymptomatic carriers i.e. $p=1$, we recover the $\mathcal{R}_0$ formula found in the SIR model. As a consequence of the Van den Driessche \& Watmough Theorem \cite{van2002reproduction}, we establish the following result regarding the local stability of the DFE.

\begin{theorem}
The continuum of DFE of system \eqref{seair_model} given by $E_{0}$ in \eqref{DFE} is locally asymptotically stable if the basic reproduction number satisfies $\mathcal{R}_{0}<1$ and unstable if $\mathcal{R}_{0}>1$.
\end{theorem}

From the basic reproduction number \eqref{R0}, it is evident that asymptomatic carriers may play an important role in the spread of COVID-19 within a population depending on their ability to transmit the infection ($\beta_A$) and their frequency in comparison with symptomatic carriers ($p$). Early estimates at the beginning of the pandemic suggested that approximately 80\% of infections were asymptomatic \cite{buitrago2020occurrence}. More recent evidence suggests only between 17\% and 20\% do not present any symptoms \cite{asymptomaticCont}. The authors in \cite{asymptomaticCont} also found that the transmission risk from asymptomatic cases appeared to be lower than that of symptomatic cases, but there was considerable uncertainty in the extent of this (relative risk 0.58; 95\% CI 0.335 to 0.994). Nevertheless, asymptomatic individuals may have more contacts than symptomatic people. There are still different opinions of the true magnitude of asymptomatic infections and their impact on the pandemic \cite{qiu2021defining}. Finally, we must remark that although model \eqref{seair_model} considers asymptomatic transmission, it is still a very simplified model that ignores the existence of a pre-symptomatic stage. As remarked in \cite{he2020temporal}, infectiousness usually starts 2.5 days before symptoms onset with a high transmission rate. So besides the non-infectious exposed period and the fully asymptomatic class, some authors have also considered a class for the pre-symptomatic stage \cite{aleta2020modelling, day2020evolutionary}

\section{The impact of non-pharmaceutical interventions}
The SEAIR model \eqref{seair_model} is proposed to study the early phase of the outbreak and therefore assumes that at the start of the pandemic no interventions were applied to control the spread of SARS-CoV-2. This makes sense for the first stage of the pandemic which, except for China, was driven by imported cases. However, as levels of local transmission began to increase, sanitary emergency measures were implemented by health authorities in several countries. Given the absence of a vaccine or effective treatment against COVID-19 at the beginning of the pandemic, preventive measures pertained to non-pharmaceutical interventions (NPIs) such as wearing a mask in public, staying at home, avoiding places of mass gathering, social distancing, ventilating indoor spaces, washing hands often, etc. 

To have a full picture of the effect of NPIs, it is necessary to formulate mechanistic mathematical models that explicitly take into account the impact induced by such sanitary measures. There are different approaches to incorporate NPIs into compartmental models, see, for example, \cite{acuna2020modeling, aguiar2020modelling, aleta2020modelling, de2020seiard,  to_mask, ngon, okuonghae2020analysis,  park2020potential, saldana2020modeling, santana2020lifting, tocto2020lockdown, ullah2020modeling}. Here, we formulate a compartmental mathematical model that explicitly incorporates: (i) isolation of infectious individuals and (ii) mitigation measures that reduce the number of contacts among the individuals in the population, namely, temporary cancellation of non-essential activities, lockdown, and social distancing. One of the key tasks throughout this pandemic has been the use of mathematical models and epidemiological data to forecast excess hospital demand. Hence, we also incorporate appropriate compartments to monitor the required hospital beds, including the number of intensive care units (ICU) during the outbreak. The new model is an extension of the SEAIR model \eqref{seair_model} and is given by the following set of differential equations:

\begin{equation}
\begin{aligned}
\dot{S} &= -\epsilon(t)(\beta_{A}A + \beta_{I}I ) S,\\
\dot{E} &= \epsilon(t)(\beta_{A}A + \beta_{I}I ) S- \sigma E,\\
\dot{A} &= (1-p)\sigma E - \gamma_{A}A,\\
\dot{I} &= p\sigma E - \gamma_{I}I,\\
\dot{Q} &=\alpha \gamma_{I}I - \delta Q,\\
\dot{H} &= \delta (1-\psi)Q - \gamma_{H}H,\\
\dot{C} &= \delta \psi Q - \gamma_{C}C,\\
\dot{R} &= \gamma_{A}A + (1-\alpha)\gamma_{I}I + \gamma_{H}(1-\mu)H + \gamma_{C}(1-\mu)C,\\
\dot{D} &= \gamma_{H}\mu H + \gamma_{C}\mu C.
\end{aligned}
\label{model_NPIs}
\end{equation} 
In model \eqref{model_NPIs}, after infection, a fraction $\alpha$ of the symptomatic individuals develop severe symptoms and are therefore isolated entering the home quarantine compartment $Q$, the other fraction recovers from the disease and enter the immune class $R$. The model assumes that once isolated, infectious individuals no longer contribute to the force of infection. Individuals expend an average of $1/\delta$ days in the home quarantine class and then a fraction $\psi$ of them develop symptoms that require hospitalization $H$ or an intensive care unit $C$. The average time individuals expend in the classes $H$ and $C$ are $1/\gamma_H$ and $1/\gamma_C$, respectively, and a fraction $\mu$ of such individuals experiment COVID-19 induced death. The function $0\leq \epsilon(t) \leq 1$ measured the reduction at a specific time $t$ in the transmission rates achieved by the implementation of a lockdown or any social distancing measures that reduce the number of contact among the population. 

\begin{figure}[hbtp]
 \centering{
    \includegraphics[width=0.5\textwidth]{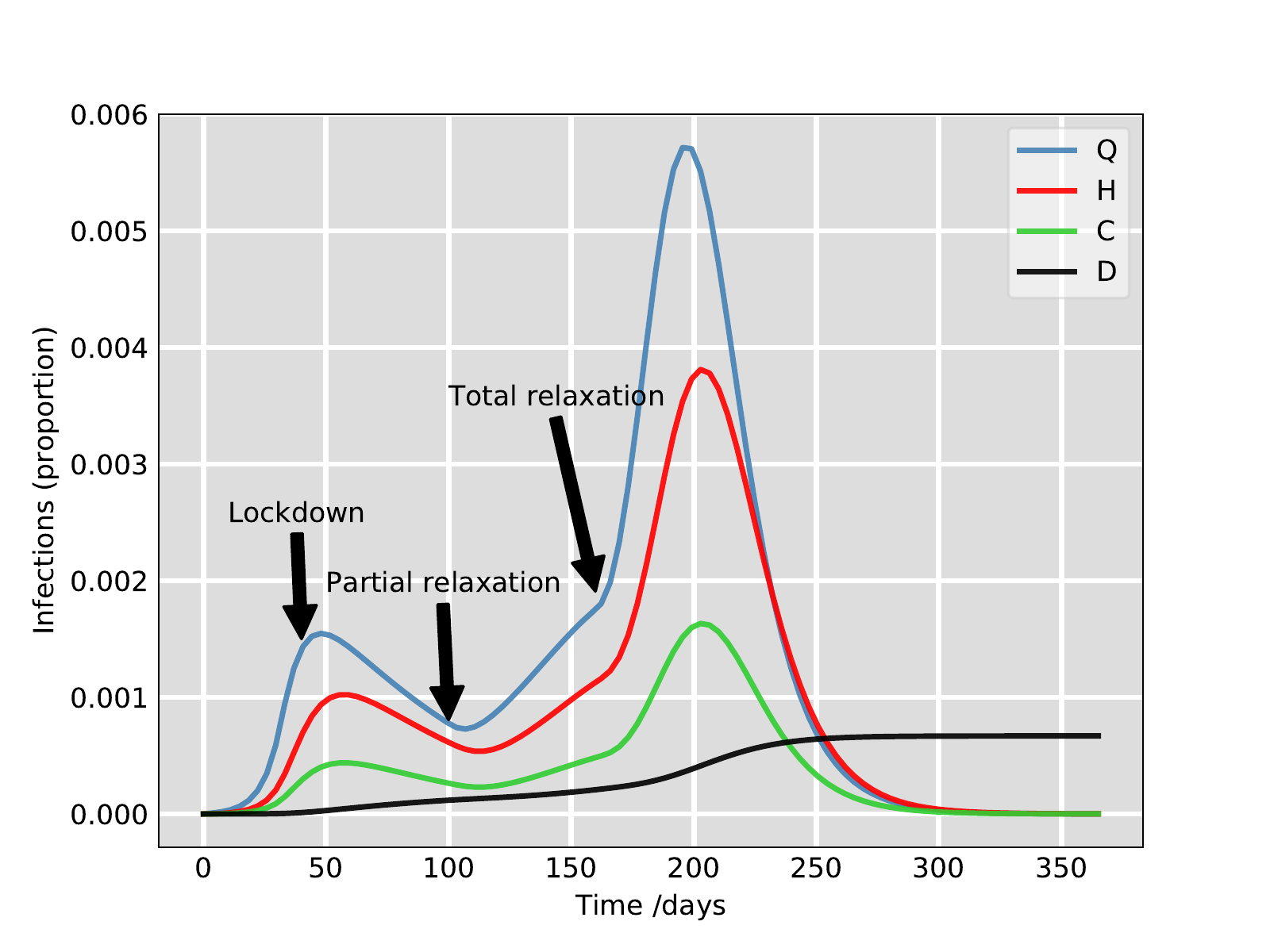}
    }
 \caption{Evolution of the daily number of home quarantine infected individuals $Q$, hospitalizations $H$, ICU occupancy $C$, and cumulative deaths $D$ depending on the relaxation of lockdowns and social distancing (isolation of infected individuals is not relaxed). Epidemiological parameters are fixed to mimic COVID-19 dynamics as follows: $\beta_{I}=0.5$ days$^{-1}$, $\beta_{A}=0.4$ days$^{-1}$, $1/\sigma=5$ days, $p=0.7$, $1/\gamma_{I} =10$ days, $1/\gamma_{A} =14$ days, $\alpha=0.1$, $\delta=1/7$ days, $\psi=0.3$, $1/\gamma_{H}=1/\gamma_{C}=7$ days, $\mu=0.01$.}
 \label{fig:npis}
\end{figure}

In Figure \ref{fig:npis}, we performed numerical simulations aimed to show how the implementation and relaxation of lockdowns impacted the transmission dynamics of COVID-19. For illustration purposes, we are considering a rather simple form for the function $\epsilon(t)$. We assume that for the first 40 days after the emergence of SARS-CoV-2 in the population no mitigation measures are implemented, so $\epsilon(t)=1$ for $0\leq t<40$. After this, health authorities implement a very strict lockdown, for approximately two months, that reduces significantly (90\% in our simulations) the number of contact within the population, thus $\epsilon=0.1$ for $30\leq t< 100$. The reduction in the prevalence of the infection due to lockdown implementation is very clear (see Figure \ref{fig:npis}). After this period, there is a partial relaxation of the lockdown to reflect the need for reactivating social and economical activities, so $\epsilon(t)=0.5$ for $100\leq t\leq 160$. This causes a new increment in the number of new infections and the start of the second wave of the epidemic (see Figure \ref{fig:npis}). Finally, after two months of partial relaxation, if health authorities decided to completely lift social distancing measures as we assume i.e. $\epsilon(t)=1$ for $t>160$, the result is an exponential increase in the prevalence of the infection that may increase healthcare system pressure.

\subsection{Optimization of non-pharmaceutical interventions}

The effectiveness of non-pharmaceutical interventions to control the epidemic has been an important aim of recent work  \cite{bairagi2020controlling, bugalia2020mathematical,  santamaria2020possible, xing2020impact}. The economic and social cost of lockdowns, bans of public events or closures of restaurants, commercial centers, etc., must be limited to reduce economic costs. Designing optimal transitory NPIs that reduce disease spread at the lower cost is a key issue that has been investigated in \cite{angulo2020simple} using a simple SIR model. Following these authors,  the design of adequate NPIs involve a  trade-off between minimizing the economic cost of their implementation and the reduction or minimization of deaths due to insufficient health support. In mathematical terms, they assume that NPIs may reduce the effective contact rate $\beta$ by a control term $(1-u)$, $u\in [0,1]$. Note that $u=1$ implies an effective contact rate of zero. However, such aim is unrealistic and thus an upper bound for $u$, defined as $u_{max}\in(0,1)$, is postulated; hence an admissible control must now satisfy $u\in[0, u_{max}]$. In this approach the control is acting on the nonlinear term of the SIR model. The authors' also assume that health services can adequately manage up to a maximum prevalence $I_{max}\in [0,1)$. This implies that the action of NPIs measured by $u$ must maintain disease prevalence below $I_{max}$. The model is governed by the following equations:
\begin{equation}
\begin{aligned}\label{tulio}
\dot{S} &= -(1-u)\beta SI,\\
\dot{I} &= (1-u)\beta SI - \gamma I,\\
\dot{R} &= \gamma I.
\end{aligned}
\end{equation}
Since it is assumed that $S(t)+I(t)+R(t)=1$ the problem is reduced to the $S-I$ plane. If the optimal NPIs problem has a solution $u^{*}$, then $u^{*}(S,I)$ gives the optimal reduction in the effective contact rate that the NPIs should achieve given that the epidemic is in the state $(S,I)$. The main result of Angulo et al \cite{angulo2020simple} is a complete analytical characterization of the optimal NPIs in the SIR model \eqref{tulio}. Their analysis shows that the solution to the optimal intervention is fully characterized by the separating curve
\begin{equation*}
 \Phi(S)= 
 \left\{ \begin{array}{lc}
 I_{max}+\mathcal{R}_{c}^{-1}In(S/S^{*})-(S-S^{*}) &  if\quad   S \geq S^{*} \\    
 \\I_{max} &  if\quad   \mathcal{R}_{0}^{-1} \leq S \leq S^{*} 
  \end{array}
  \right.   
\end{equation*}
where $S^{*} = \min \left\lbrace \mathcal{R}_{c}, 1 \right\rbrace$, and $\mathcal{R}_{c}=(1-u_{max})\mathcal{R}_{0}$ is a controlled reproduction number. Note that the shape of the separating curve depends on $I_{max}$ and $\mathcal{R}_c$. If $\mathcal{R}_{c}\leq 1$, the separating curve is the straight line $\Phi(S) = I_{max}$. When $\mathcal{R}_c > 1$, the separating curve becomes nonlinear. The optimal control intervention is characterized by the separating curve as follows:
\begin{itemize}
    \item[(1)] an optimal intervention exists if and only if the initial state $(S_0, I_0)$ lies below this separating curve, that is, $I_0\leq \Phi(S_0)$
    \item[(2)] if it exist, the optimal intervention $u^{*}$ takes the feedback form 
    \begin{equation*}
         u^{*}(S,I)= 
         \left\{ \begin{array}{lcc}
         0 &  if\quad I<\Phi(S)\; or\; S\leq \mathcal{R}_{0}^{-1} \\    
          \\u_{max} &  if\quad  I=\Phi(S)\; and\; S\geq S^{*}\\
          \\ 1-1/(\mathcal{R}_{0}S) &  if\quad I=\Phi(S)\; and\; S^{*}\geq S>\mathcal{R}_{0}^{-1} \\
  \end{array}
  \right.   
    \end{equation*}
\end{itemize}
Hence, the optimal intervention starts when $I(t)$ reaches $\Phi(S(t))$, and then it slides $I(t)$ along $\Phi(S)$ until reaching the region where $S\leq \mathcal{R}_{0}^{-1}$. Observe that once the trajectory reaches this region, the prevalence of the infection will decrease without the need of further control \cite{angulo2020simple}.

\section{Vaccination policies, herd-immunity and the effective reproduction number}\label{sec:vac}

Besides being an indicator of the severity of an epidemic, the basic reproduction number $\mathcal{R}_0$ is a powerful tool to estimate the control effort needed to eradicate a disease. Consider an infection in a population that mixes homogeneously with $\mathcal{R}_{0}>1$. If at least $\mathcal{R}_{0}-1$ of these individuals are protected from the infection either through naturally acquired immunity after infection or through vaccine-induced immunity, then the epidemic cannot grow \cite{keeling2013mathematics}. Hence, the infection can be eradicated if a fraction of individuals greater than
\begin{equation}
q=(\mathcal{R}_0 -1)/\mathcal{R}_0 = 1- 1/\mathcal{R}_0 
\end{equation}
has been afforded lifelong protection. Under these conditions, a considerable fraction of the population is immune providing indirect protection to those still susceptible. Hence, on average, a typical infectious individual no longer generates more than one secondary infection and herd immunity is established. Estimations of the basic reproduction number for SARS-CoV-2 usually range between $2.0-4.5$, so for example, if $\mathcal{R}_0 = 2.5$ as estimated for Mexico by \cite{saldana2020modeling} (although there are several other estimates for this parameter e.g., \cite{acuna2020modeling,mena2020using}) a vaccination program aiming to attain herd immunity must immunize at least 60\% of the population. This result is only valid for a perfect vaccine that prevents infection with 100\% efficacy; however, for most infections, vaccines only confer partial protection. For the case of SARS-CoV-2 infection, vaccine developers behind the Pfizer/BioNTech, Moderna, and Gamaleya vaccines have announced that their vaccines have efficacy above $90\%$, but other developers such as the AstraZeneca and Sinovac teams have vaccines with efficacy way below 90\% \cite{kim2021looking}. For vaccines with efficacy $0<\psi<1$ ($\psi=1$ means 100\% efficacy), if health authorities vaccine a fraction $v$ of the population and the remaining fraction of the population are susceptible $s$, then $R_v = s\mathcal{R}_{0} + (1-\psi)v\mathcal{R}_{0}$ is the mean number of infections generated by a typical infectious person in this partially immunized population. So the minimum level of vaccination needed to eradicate the infection (the minimum $v$ such that $R_v<1$) becomes $v=q/\psi=(1-1/\mathcal{R}_{0})/\psi$.

In the absence of mitigation measures, herd immunity is achieved when the effective reproduction number $\mathcal{R}_e$ (also denoted $\mathcal{R}_t$) is equal to unity which in turn corresponds to the time when the peak of the epidemic is reached \cite{weitz2020awareness}. The effective reproduction number $\mathcal{R}_e$ quantifies the mean number of infections produced by a typical infectious case in a partially immune or protected (via isolation, quarantine, etc) population \cite{delamater2019complexity}. Unlike, $\mathcal{R}_0$, the effective reproduction number does not assume a fully susceptible population and changes over time depending on the population's immune status and the impact of non-pharmaceutical interventions or vaccination programs in the mitigation of further disease transmission. Under the classical homogeneous mixing assumption, the effective reproduction number at a particular time $t$ can be approximated as 
\begin{equation}
\mathcal{R}_e(t) \approx (1-p_{c}(t))\mathcal{R}_0 S(t)/N(t),\label{Re}
\end{equation}
where $p_{c}(t)$ is the reduction in the transmission rate due to mitigation measures where the effectiveness of such measures may vary at a particular time. The depletion of the susceptible pool decreases the value of $\mathcal{R}_e$. Observe that without mitigation measures ($p_c=0$), at the beginning of the outbreak, $\mathcal{R}_e(0) = \mathcal{R}_0$.

The time-dependent effective reproduction number $\mathcal{R}_{e}(t)$ has been of paramount importance to assess the impact of mitigation measures against COVID-19 and to guide the easing of such restrictions. There are several methodologies to estimate the value of $\mathcal{R}_e$ from data, see, for example \cite{bettencourt2008real, cori2013new, contreras2020real, nishiura2009effective,  wallinga2004different}. Probably, the most simple mathematical formulation to obtain the time-dependent effective reproduction number directly from epidemiological data is the method recently proposed by Contreras and coworkers \cite{contreras2020real}. This method considers that, in real-life situations, during an epidemic outbreak, the effective contact rate $\beta$ is not constant as assumed in \eqref{sir}. Instead, it is plausible to assume that $\beta(t)$ is a function that varies in time depending on several circumstances, for example, the impact of mobility restrictions and other mitigation measures to control disease's spread. Then, the effective reproduction number is
\begin{equation}
\mathcal{R}_e (t)=\dfrac{\beta(t)}{\gamma}\dfrac{S(t)}{N(t)}.
\end{equation}
Considering the SIR model equations \eqref{sir} with a time-dependent $\beta(t)$, a direct computation using the chain rule allow us to obtain \cite{contreras2020real}
\begin{equation}
\dfrac{dI}{dS}= -1 +\dfrac{1}{\mathcal{R}_{e}(t)}.\label{chain_rule}
\end{equation}
The equation \eqref{chain_rule} is discretized in an interval $[t_{i-1}, t_{i}]$ where it is assumed that $\mathcal{R}_{e}(t)=\mathcal{R}_{e}(t_i)$ is constant, thus
\begin{equation}
\mathcal{R}_{e}(t_i) = \dfrac{1}{\dfrac{\Delta_i I}{\Delta_i S} + 1}
\end{equation}
An extension of the SIR model considering the number of deaths and a population balance implies that discrete differences satisfy $\Delta_i S + \Delta_i I + \Delta_i r + \Delta_i D = 0 $, hence
\begin{equation}
\mathcal{R}_{e}(t_i) = \dfrac{\Delta_i I}{\Delta_i R + \Delta_i D} + 1.\label{last_re}
\end{equation}
As remarked in \citep{contreras2020real}, if we note that $\Delta_i I = (-\Delta_i S) - (\Delta_i R + \Delta_i D)$, that is, the change in the prevalence, $\Delta_i I$, equals new infections, $-\Delta_i S$, minus new recoveries plus new deaths, $(\Delta_i R + \Delta_i D)$, therefore
\begin{equation}
\mathcal{R}_{e}(t_i) = \dfrac{-\Delta_i S}{\Delta_i R + \Delta_i D}=\dfrac{\textrm{New infections} }{\textrm{New recoveries} + \textrm{New deaths} }.
\end{equation}
The features, advantages, limitations, and practical application of this methodology to obtain $\mathcal{R}_e$ for COVID-19 is studied in more detail in \cite{contreras2020real, medina2020country}

\subsection{ COVID-19 vaccine prioritization}

Deployment of COVID-19 vaccines was initiated in several countries at the beginning of January 2021.  As expected, there has been a high demand for the limited supplies of COVID-19 vaccines, so the optimization of vaccine allocation to maximize public health benefit has been a problem of interest \cite{Fitzpatrick2021}.
Mathematical models have also been used to guide public health policies on the optimization of vaccine allocation \cite{bubar2021, buckner2021dynamic, castro2021prioritizing, goldstein2021vaccinating, schmidt2020covid, Fitzpatrick2021}. It is well known that a vaccine directly protects those vaccinated but also indirectly protects those that are not, the more efficacious is the vaccine the greater and more beneficial the indirect protection will be. Recently, Bubar et al.  \cite{bubar2021} developed a suite of mathematical models geared to evaluate age-specific vaccine prioritization policies. Several of the vaccines currently available have high levels of efficacy (above 90\%) which means that they can be effective in blocking transmission. On the other hand, influenza is a well-known respiratory infection for which vaccination policies do exist and may therefore serve as an important reference to those for SARS-CoV-2. However, the age-specific probability of infection and the age-specific mortality are different in these two diseases with a higher burden for older than 60 years old people \cite{Fitzpatrick2021}.

The model used by Bubar et al \cite{bubar2021} is an age-structured SEIR model with a force of infection, $\lambda_i$, for a susceptible individual in age group $i$ given as follows:
\begin{equation*}
\lambda_i=\phi_i\sum_{j=1}^nc_{ij}\frac{I_j+I_{vj}+I_{xj}}{N_j-D_j}
\end{equation*}
where $\phi_{i}$ is the probability of a successful transmission given contact with an infectious individual, $c_{ij}$ is the daily contact rate of individuals in age group $i$ with individuals in age group $j$, $I_j$ is the number of infectious unvaccinated individuals, $I_{vj}$ are individuals who are vaccinated yet infectious, and $I_{xj}$ is the number of infectious individuals that are ineligible for vaccination for the personal hesitancy of due to a positive serological test, $N_j$ is the number of individuals in the age group $j$ and $D_j$ are the individuals of age group $j$ who have died.

Bubar et al \cite{bubar2021} incorporated vaccine hesitancy assuming limited vaccine uptake such that at most 70\% of any age group was eligible to be vaccinated. Such constrain was performed assuming that 30\% of each class for each age group was initialized as ineligible for vaccination. Moreover, in their more parsimonious model, they assumed the vaccine to be both, transmission- and infection- blocking, and to work with variable efficacy. In particular, they considered two ways to implement vaccine efficacy ($ve$): as an all-or-nothing vaccine, where the vaccine yields perfect protection to a fraction $ve$ of people who receive it, or as a leaky vaccine, where all vaccinated people have reduced probability $ve$ of infection after vaccination. To incorporate age-dependent vaccine efficacy, they parameterized the relationship between age and vaccine efficacy via an age-efficacy curve with (i) a baseline efficacy, an age at which efficacy begins to
decrease (hinge age), and a minimum vaccine efficacy $ve_m$ for adults $80+$. This was done assuming that $ve$ is equal to a baseline value for all ages younger than the hinge age, then decreases step-wise in equal increments for each decade to the specified minimum $ve_m$ for the $80+$ age group. Finally, existing seroprevalence estimates were Incorporated varying the basic reproduction number and the percentage cumulative incidence reached. 

The model outcomes used to compare the performance of different vaccine allocation strategies were the cumulative number of infections, deaths, and years of life lost. Bubar et al \cite{bubar2021} concluded that although vaccination targeted to younger people (20-50 years old) minimized cumulative incidence, mortality and years of life lost were minimized when applied first to older people. These results were based on numerical simulations with a time horizon of one year after the date of the vaccine introduction arguing that this allowed them to focus on the early prioritization phase of the COVID-19 vaccination programs.

\section{Superspreading events}

In the beginning of the epidemic, surveillance made extensive use of the basic reproduction number $\mathcal{R}_{0}$ to characterize the average number of secondary infections. Nevertheless, $\mathcal{R}_{0}$ may hide a large variation at the individual level. Indeed, after more than one year into the pandemic, there have been several reports of superspreading events (SSEs) in which many individuals are infected at once by one or few infectious carriers (see, for example, \cite{endo2020estimating, kochanczyk2020super, lemieux2020phylogenetic, liu2020secondary, ndairou2020mathematical, santana2020lifting} and the references therein). Current reports suggest that a small group of infections generate most of secondary cases\cite{kain2021chopping}. In other words, even if $\mathcal{R}_{0}\approx 2-3$, most individuals are not infecting 2 or 3 other people; instead, a tiny number of people dominate transmission while an average person do not transmit the virus at all \cite{althouse2020superspreading}. Current data suggest that prolonged indoor gatherings with poor ventilation are one of the main factors inducing SSEs. Hence, crowded closed places are hotspots for SSEs and a source of COVID-19 infections. More factors may lead to SSEs. For example, events in which a huge number of people temporarily cluster, so the number of contacts suddenly increases far above their mean \cite{althouse2020superspreading}. The existence of superspreaders, that is, individuals having biological features that cause them to shed more virus than others, is another issue that has attracted attention \cite{nature_superspreading}.    

To understand the role of individual variation in outbreak dynamics, Lloyd-Smith et al \cite{lloyd2005superspreading}, introduced the individual reproduction number, $\nu$, as a random variable that assess the average number of secondary cases generated by a particular infected individual. Then $\nu$ values can be estimated from a continuous probability distribution with a population mean $\mathcal{R}_0$. In this context, SSEs are realizations from the right-hand tail of a distribution of $\nu$. A Poisson process is used to describe the stochastic essence of the transmission process; hence, the expected number of secondary infected generated by each case, $Z$, is approximated by an offspring distribution $\mathbb{P}(Z = k)$ where $Z\sim Poisson(\nu)$. Lloyd-Smith et al. considered three possible distributions of $\nu$ generating three candidate models for the offspring distribution:
\begin{itemize}
    \item[(1)] Generation-based models neglecting individual variation, i.e. $\nu=\mathcal{R}_0$ for all cases, yielding $Z\sim Poisson(\mathcal{R}_0)$.
    \item[(2)] Differential-equation models with
     homogeneous mixing transmission and constant recovery rates, $\nu$ is exponentially distributed yielding \\$Z\sim geometric(\mathcal{R}_{0})$.
     \item[(3)] $\nu$ is gamma-distributed with mean $\mathcal{R}_0$ and dispersion parameter $k$, yielding $Z \sim NB (\mathcal{R}_{0},k)$ (NB$=$ Negative Binomial).
\end{itemize}

Observe that the NB model includes the Poisson ($k\rightarrow \infty$) and geometric ($k= 1$) models as special cases. It has variance $\mathcal{R}_{0}(1 + \mathcal{R}_{0}/k)$, so smaller values of $k$ indicate greater heterogeneity. Althouse et al \cite{althouse2020superspreading} showed that an epidemic outbreak dominated by SSEs and an NB distribution of secondary infections with small $k$ has very different transmission dynamics in comparison to a Poisson model outbreak with the same $\mathcal{R}_0$. For the NB model, secondary infections are over-dispersed causing early transmission dynamics that are more stochastic. Hence, in this case, an epidemic outbreak has lower probability to grow into a huge epidemic. Nevertheless, if the outbreak takes off under the NB model, the incidence starts showing stable exponential growth, with a growth rate approaching that of a model with the same $\mathcal{R}_0$, but a Poisson distribution of secondary infections
i.e., an NB model with $k\rightarrow \infty$. However, during the early phase of an NB outbreak that takes off, disease incidence will look more intense in the first few generations when SSEs will generate the most of secondary infections, making it possible to spin out large infection clusters in few generations, whereas a Poisson model cannot \cite{althouse2020superspreading, lloyd2005superspreading}.

In the context of COVID-19, Endo et al \cite{endo2020estimating} presented the first study (to the authors' knowledge) estimating the level of overdispersion in COVID-19 transmission by using a mathematical model that is characterized by $\mathcal{R}_0$ and the overdispersion parameter $k$ of a NB branching process. Their results suggest a high degree of individual-level variation in the transmission of COVID-19. Assuming that the $\mathcal{R}_0$ lies between $2-3$, Endo et al. estimated an overdispersion parameter $k$ to be around $0.1$ (median estimate $0.1$; 95\% CrI: $0.05-0.2$ for $\mathcal{R}_0 = 2.5$), suggesting that 80\% of secondary transmissions may have been caused by a small fraction of infectious individuals (~10\%). Other studies have also estimated $k$ values and their results suggest that $k$ lies in the range $0.04-0.3$ \cite{hebert2020beyond, liu2020secondary, kain2021chopping}.

\section{Viral evolution and vaccine escape mutants}
As of April 2021, several variants of the SARS-CoV-2 have been reported globally \cite{cdc, chen2021covid, mercatelli2020geographic, phan2020genetic}. RNA viruses, such as the coronavirus SARS-CoV-2, will naturally mutate over time so such variants are not unexpected. Moreover, most mutations are irrelevant in an epidemiological context. Nevertheless, of the multitude of variants circulating worldwide, at the time of writing (April 2021), health experts are mainly worried about three variants that have undergone changes to the spike protein and are maybe more infectious and threatening \cite{fontanet2021sars}. The first of them is the B.1.1.7 variant that was first identified in the United Kingdom (UK) and seems to be more transmissible than other variants currently circulating. This variant has also been linked with an increased risk of death but there is still uncertainty surrounding this result. The South-African variant B.1.351 emerged independently from B.1.1.7 but share some of its mutations. The third is the Brazilian variant P.1 that contains some additional mutations that and may be able to overcome the immunity developed after infection by other variants \cite{cdc}.

Current COVID-19 vaccines were developed before the emergence of the above-mentioned variants. Hence, another major concern is the possibility that variants will make vaccines less effective. Some preliminary results by leading vaccine developers suggest that vaccines can still protect against the new variants \cite{fontanet2021sars}. However, the vaccine-induced immune response may not be as strong or long-lasting \cite{bbc}. Apart from these problems, the initial and limited vaccine supply has raised some discussion on how to distribute COVID-19 vaccines \cite{castro2021prioritizing, de2021se, goldstein2021vaccinating, iboi2020will, makhoul2020epidemiological, saldana2020trade}. Beyond prioritizing healthcare workers and the elderly, the optimal strategy for the general public remains complex. Some countries, including Canada and the UK, had proposed to delay the second dose of the vaccine as an attempt to increase the number of individuals receiving at least one dose and therefore gaining more protection within the population. Delaying the second dose could create conditions that promote the evolution of vaccine escape, namely, viral variants resistant to the antibodies created in response to vaccination \cite{saad2021epidemiological}. Escape variants have the potential of creating more infections, deaths and prolong the pandemic. In general, the start of vaccination programs all around the world, whilst the pandemic is still ongoing, may rapidly exert selection pressure on the SARS-CoV-2 virus and lead to mutations that escape the vaccine-induced immune response \cite{gog2021vaccine}. At this time, there is uncertainty around the strength of such selection and the probability of vaccine escape, so more studies are needed in this direction. Recently, Gog et al \cite{gog2021vaccine} studied how considerations of vaccine escape risk might modulate optimal vaccine priority order. They found two main insights: (i) vaccination aimed at reducing prevalence could be more effective at reducing disease than directly vaccinating the vulnerable; (ii) the highest risk for vaccine escape can occur at intermediate levels of vaccination. They also remarked that vaccinating most of the vulnerable and only a few of the low-risk individuals could be extremely risky for vaccine escape. Their results are based on a two-population model with differing vulnerability and contact rates.

The existence of geographic regions of the human population where the vaccine is scarce is another concern. It can be argued that regions that do not have access to the vaccine can serve as evolutionary reservoirs from which vaccine escape mutations may arise. This hypothesis has been explored by Gerrish et al \cite{Gerrish2021} using a simple two-patch deterministic epidemic model as follows. They considered COVID-19 epidemics in two neighboring regions or patches. Assuming that only one patch has access to the vaccine, they investigated if the presence of the unvaccinated patch affects the probability of vaccine escape in the vaccinated one. 

The model follows the SIR structure for both patches augmented to consider vaccine escape mutants. The governing equations are as follows: 

\begin{equation}\label{eq:epi1}
\begin{gathered}
\dot{S_{j}} =  - \sum_k^n \beta_{kj} I_{k} S_{j} - \phi_{j} S_{j}, ~~~~~~~
\dot{I_{j}} = \sum_k^n \beta_{jk} I_{j} S_{k}  - (\gamma + \mu) I_{j}, \\
\dot{V_{j}} = \phi_{j} S_{j}, ~~~~~
\dot{E_{j}} = \mu I_{j}  -  \gamma E_{j}, ~~~~~
\dot{R_{j}} = \gamma (I_{j} + E_{j}),
\end{gathered}
\end{equation}
where $S_{j}$, $V_{j}$, $I_{j}$, $E_{j}$, and $R_{j}$ are the fraction of the population that are susceptible, vaccinated, infected, infected with escape mutant, and recovered, respectively, in patch $j$; $\beta_{ij}$ is the transmission rate from patch $i$ to patch $j$, $\phi_j$ is vaccination rate in patch $j$; $\gamma$ is recovery rate; $\mu$ is a composite per-host mutation rate from wildtype virus to escape mutant virus; $n$ is the number of patches but they explored only the simple two-patches case ($n=2$).

Escape mutants are not considered explicitly, because, under normal conditions, they will always have a strong selective advantage in a vaccinated population. Thus the model focuses on the timing of the first infection event in which a new host is infected with an escape mutant.

The time of the first infection event in which a new host in patch $j$ is infected by an escape mutant that arose in patch $i$ will occur at a rate $r_{ij}(t)=\beta_{ij}E_{i}(t)(S_{j}(t)+\sigma V_{j}(t))$, where $\sigma$ allows for varying levels of escape and ranges from $\sigma=0$, no escape mutations, to full escape $\sigma=1$ \cite{Gerrish2021}.  It is also assumed that intra-patch transmission rates are equal, $\beta_{jj}=\beta$, and inter-patch transmission rates are equal too, $\beta_{ij}\big|_{i\neq j} =\beta_0$. Assuming $\beta_{0} <\beta$ i.e.,  inter-patch transmission is  much less frequent than intra-patch transmission, the random variable $T_f$ is defined as the time at which the last infected individual recovers. Gerrish et al \cite{Gerrish2021} showed that the probability of vaccine escape can be orders of magnitude higher if vaccination is fully allocated to one patch and no vaccination allocated to the other. On the contrary,  equal vaccination on the two patches gives the lowest probability of vaccine escape. In other words, unequal distribution of COVID-19 vaccines may lead to vaccine escape and hence, support arguments for vaccine equity at all scales.

\section{Conclusions}
%
%
Once the pandemic potential of the SARS-CoV-2 became established in early 2020, there have been many modeling efforts by scientific and public health communities to improve our understanding of the underlying mechanism of SARS-CoV-2 transmission and control \cite{thompson2020key}. Mathematical and computational models have been key tools in the fight against the COVID-19 pandemic helping to forecast hospital demand, evaluate the impact of non-pharmaceutical interventions, guide lockdown exit strategies, the potential reach of herd-immunity levels, evaluate the optimal allocation of COVID-19 vaccines, and in general to provide guidance to policy-makers about the development of the epidemic \cite{aguiar2020modelling, aleta2020modelling, castro2021prioritizing, contreras2020real, eker2020validity, goldstein2021vaccinating, makhoul2020epidemiological,medina2020country, mena2020using, ngon, saldana2020modeling, saldana2020trade, santana2020lifting,  weitz2020awareness}. 

In this work, we have discussed the role of the so-called compartmental models to gain insight into epidemiological aspects of virus transmission and the possibilities for its control. The main objective has been to present an overview of the mathematical methods behind some important aspects of the general theory of infectious disease modeling. We started revisiting simple deterministic compartmental models closely connected with the classical work of Kermack \& McKendrick and important epidemiological quantities such as the basic and effective reproduction numbers. Then, we explored the role of the so-called mixing function which allows introducing heterogeneity in epidemic models and is, therefore, a key component in disease modeling. Then we investigated a number of compartmental epidemic models to study the early phase of the COVID-19 epidemic outbreak, how awareness-driven behavior modulates the epidemic shape and the role of asymptomatic carriers in disease transmission. The impact of non-pharmaceutical interventions for COVID-19 control is also discussed. We also revisited important results related to vaccination policies, herd-immunity, and the effective reproduction number together with a simple method to perform real-time estimation of this last quantity. We also discussed the presence of superspreading events in the pandemic, the possibility of viral evolution and vaccine escape mutants.

One important topic that is not discussed in this review is the interaction of COVID-19 with other respiratory illnesses. Co-circulation of SARS-CoV-2 and other endemic respiratory viral infections is a potential reality that can bring more challenges to public health. Currently, there have been some concerns about the interaction between SARS-CoV-2 and influenza viruses and preliminary results suggest that an initial infection with the influenza A virus strongly enhances the infectivity of SARS-CoV-2 \cite{bai2021coinfection}. Some studies have also already reported proportions of SARS-CoV-2 co-infections with other respiratory viruses \cite{burrel2021co}. Co-infection mechanisms are common in nature and the previously mentioned studies highlight the risk of influenza virus and SARS-CoV-2 co-infection to public health. Tailoring epidemic models to incorporate the most important features of COVID-19 such as the existence of asymptomatic carriers, the use of non-pharmaceutical interventions, the recent introduction of vaccinations, and the emergence of new variants of SARS-CoV-2 together with co-circulation with other pathogens will result in very complex dynamics. Therefore, developing realistic mathematical models to study the co-circulation of SARS-CoV-2 and other respiratory infections presents one important challenge for disease modelers \cite{zegarra2021co}. There are other topics that are not treated in this review. For example, simulation models for the spatial spread of SARS-CoV-2, age and risk structure, the role of human behavior on disease dynamics, parameter estimation techniques to calibrate models with official data, exploration of long-term epidemiological outcomes such as the possibility of recurrent seasonal outbreaks, among others. Nevertheless, we believe we have given at least a brief overview of key modeling efforts and current challenges related to COVID-19.

The immense number of publications related to the ongoing COVID-19 pandemic has confirmed a fundamental fact: the strategic use of mathematical modeling in public health is a multidisciplinary activity that requires a critical judgment in the interpretation of the underlying model's assumptions and their impact on the projections and outcomes. The use of mathematical models to evaluate contingency plans is essential to overcome a public health emergency. However, a considerable effort is still needed to improve the credibility and usefulness of epidemiological so we are better to prepare to respond to future epidemics. 

\section*{Acknowledgments}
We acknowledge support from DGAPA-PAPIIT-UNAM grants IV100220 (proyecto especial COVID-19) and IN115720.


\bibliographystyle{apalike}
\bibliography{references}



\end{document}